\documentclass[aps,prl,preprintnumbers,twocolumn,groupedaddress,nofootinbib]{revtex4}

\usepackage{latexsym}
\usepackage{amsmath}
\usepackage{amsfonts}
\usepackage{graphicx}

\usepackage{color}

\allowdisplaybreaks

\newcommand{\bq}{\begin{eqnarray}}
\newcommand{\eq}{\end{eqnarray}}
\newcommand{\eps}{\varepsilon}

\newcommand{\arxivdate}{April 30, 2018}

\newcommand{\curveone}{(a)}
\newcommand{\curvetwo}{(b)}
\newcommand{\curvethree}{(c)}

\begin{document}

\preprint{MITP/18-033}
\title{\boldmath{The planar double box integral for top pair production with a closed top loop to all orders in the dimensional regularisation parameter}}

\author{Luise Adams, Ekta Chaubey and Stefan Weinzierl}
\affiliation{PRISMA Cluster of Excellence, Institut f{\"u}r Physik, Johannes Gutenberg-Universit\"at Mainz, D-55099 Mainz, Germany}

\date{\arxivdate}

\begin{abstract}
We compute systematically 
for the planar double box Feynman integral relevant to top pair production with a closed top loop
the Laurent expansion in the dimensional regularisation parameter $\varepsilon$.
This is done by transforming the system of differential equations for this integral and all its sub-topologies to
a form linear in $\varepsilon$, where the $\varepsilon^0$-part is strictly lower triangular.
This system is easily solved order by order in the dimensional regularisation parameter $\varepsilon$.
This is an example of an elliptic multi-scale integral involving several elliptic sub-topologies.
Our methods are applicable to similar problems.
\end{abstract}

\maketitle

\section{Introduction}
\label{sec:intro}

The physics of heavy elementary particles like the Higgs boson, the top quark or the $W$- and $Z$-bosons
plays an important role at the LHC and future colliders.
Precision particle physics at these colliders
relies crucially on our abilities to perform higher-order perturbative calculations
and in particular on our abilities to compute the relevant Feynman integrals.
The method of differential equations \cite{Kotikov:1990kg,Kotikov:1991pm,Remiddi:1997ny,Gehrmann:1999as,Argeri:2007up,MullerStach:2012mp,Henn:2013pwa,Henn:2014qga,Adams:2017tga}
has been used successfully for many Feynman integrals
which evaluate to multiple polylogarithms \cite{Goncharov_no_note,Goncharov:2001,Borwein,Moch:2001zr,Vollinga:2004sn}.
For a large number of scattering processes with massless particles this is sufficient.
However, as soon as massive particles enter the game, it is known that starting at two loops
multiple polylogarithms will not be sufficient to express the Feynman integrals.
The simplest example of a Feynman integral not expressible in terms of multiple polylogarithms
is the two-loop sunrise integral with equal non-zero internal masses \cite{Broadhurst:1993mw,Berends:1993ee,Bauberger:1994nk,Bauberger:1994by,Bauberger:1994hx,Caffo:1998du,Laporta:2004rb,Kniehl:2005bc,Groote:2005ay,Groote:2012pa,Bailey:2008ib,MullerStach:2011ru,Adams:2013nia,Bloch:2013tra,Adams:2014vja,Adams:2015gva,Adams:2015ydq,Remiddi:2013joa,Bloch:2016izu,Groote:2018rpb}.
This integral is related to an elliptic curve and can be expressed to all orders in the 
dimensional regularisation parameter $\eps$ in iterated integrals of modular forms of $\Gamma_1(6)$.
Integrals, which do not evaluate to multiple polylogarithms are now an active field of 
studies in particle physics \cite{Bloch:2014qca,Remiddi:2016gno,Adams:2016xah,Adams:2017ejb,Bogner:2017vim,Adams:2018yfj,Sogaard:2014jla,Bonciani:2016qxi,vonManteuffel:2017hms,Primo:2017ipr,Ablinger:2017bjx,Bourjaily:2017bsb,Hidding:2017jkk,Passarino:2017EPJC,Remiddi:2017har,Broedel:2017kkb,Broedel:2017siw,Broedel:2018iwv}
and string theory \cite{Broedel:2014vla,Broedel:2015hia,Broedel:2017jdo,DHoker:2015wxz,Hohenegger:2017kqy,Broedel:2018izr}.

In this letter we report on a more involved computation.
We consider the planar double box integral relevant to top-pair production with a closed top loop.
This integral enters the next-to-next-to-leading order (NNLO) contribution for the process $pp \rightarrow t \bar{t}$.
Up to now, it is not known analytically.
The existing NNLO calculation for this process uses numerical approximations for this integral \cite{Czakon:2013goa,Baernreuther:2013caa}.
Our inability to compute this integral analytically has been a show-stopper for further progress on the analytical side.
In this letter we show how to compute analytically this integral.
Our methods are applicable to similar problems.

The planar double box integral depends on two scales (for example $s/m^2$ and $t/m^2$, where $s$ and $t$ are the usual Mandelstam variables
and $m$ the mass of the heavy particle).
It involves the sunrise graph as a sub-topology.
Therefore, we do not expect this integral to evaluate to multiple polylogarithms.
Phrased differently, we expect to see elliptic generalisations of multiple polylogarithms.
An obvious question is: Which elliptic curve?
To some surprise, there is not a single elliptic curve associated to this integral, but three different ones.
We show in this letter how to extract the elliptic curves from the maximal cuts of the (sub-) topologies.
From these elliptic curves we obtain their periods.

In the next step we bring the system of differential equations to a form linear in $\eps$,
where the $\eps^0$-part is strictly lower triangular.
We introduce kinematic variables $x$ and $y$, which rationalise the square roots in the polylogarithmic case (i.e. for $t=m^2$).
The transformation of the basis of master integrals is not rational in $x$ and $y$, however we find a transformation 
which is rational in $x$, $y$, the periods of the three elliptic curves and their $y$-derivatives.
Note that a system of differential equations linear in $\eps$, where the $\eps^0$-part is strictly lower triangular,
can easily transformed to an $\eps$-form (i.e. without any $\eps^0$-part)
by introducing primitives for the terms occurring in the $\eps^0$-part.
Both systems are equivalent and both are easily solved order by order in the dimensional regularisation parameter $\eps$.
For the case at hand the required primitives are usually transcendental functions.
We prefer to work with a system linear in $\eps$, where in the transformation matrix only the periods and their derivatives
occur as transcendental functions.

There are two interesting cases, where the solution for the Feynman integrals simplify:
for $t=m^2$ the solution degenerates to multiple polylogarithms,
for $s=\infty$ the solution degenerates to iterated integrals of modular forms for $\Gamma_1(6)$.

\section{The integral}
\label{sec:integral}

\begin{figure}
\begin{center}
\includegraphics[scale=1.0]{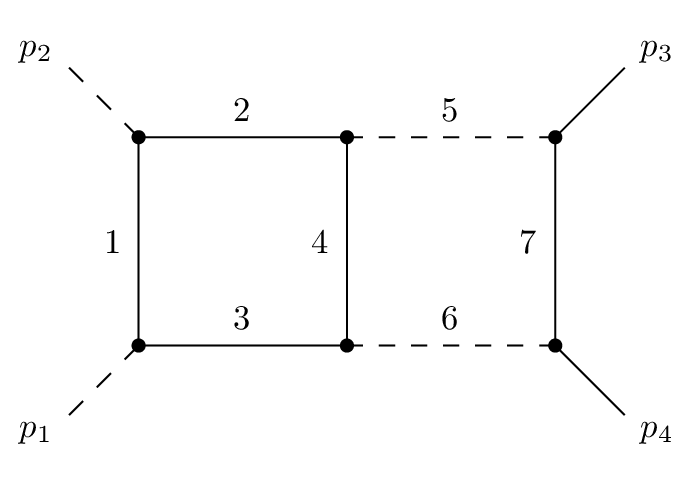}
\end{center}
\caption{
The planar double box.
Solid lines correspond to massive propagators of mass $m$, dashed lines correspond to massless propagators.
All external momenta are out-going and on-shell: $p_1^2=p_2^2=0$ and $p_3^2=p_4^2=m^2$.
}
\label{fig_double_box_graph}
\end{figure}

We consider the planar double box integral shown in fig.~(\ref{fig_double_box_graph}).
This integral is relevant to the NNLO corrections for $t\bar{t}$-production at the LHC.
In fig.~(\ref{fig_double_box_graph}) the solid lines correspond to propagators with a mass $m$,
while dashed lines correspond to massless propagators.
All external momenta are out-going and on-shell.
The Mandelstam variables are defined as usual
\bq
 s = \left(p_1+p_2\right)^2,
 & &
 t = \left(p_2+p_3\right)^2.
\eq
We are interested in the dimensional regulated integral
\bq
\label{def_integral}
\lefteqn{
 I_{\nu_1 \nu_2 \nu_3 \nu_4 \nu_5 \nu_6 \nu_7}\left( D, \frac{s}{m^2}, \frac{t}{m^2} \right)
 = } & &
 \nonumber \\ 
 & &
 e^{2 \gamma_E \eps}
 \left(m^2\right)^{\sum\limits_{j=1}^7 \nu_j - D}
 \int \frac{d^Dk_1}{i \pi^{\frac{D}{2}}} \frac{d^Dk_2}{i \pi^{\frac{D}{2}}}
 \prod\limits_{j=1}^7 \frac{1}{ P_j^{\nu_j} },
\eq
where $\gamma_E$ denotes the Euler-Mascheroni constant, 
$D=4-2\eps$ denotes the dimension of space-time
and the propagators are given by
\begin{align}
 P_1 & = -\left(k_1+p_2\right)^2 + m^2,
 &
 P_2 & = -k_1^2 + m^2,
 \nonumber \\
 P_3 & = -\left(k_1+p_1+p_2\right)^2 + m^2,
 &
 P_4 & = -\left(k_1+k_2\right)^2 + m^2,
 \nonumber \\
 P_5 & = -k_2^2,
 &
 P_6 & = -\left(k_2+p_3+p_4\right)^2,
 \nonumber \\
 P_7 & = -\left(k_2+p_3\right)^2 + m^2.
 &
\end{align}
This integral has a Laurent expansion in $\eps$:
\bq
\label{Laurent_expansion}
 I_{\nu_1 \nu_2 \nu_3 \nu_4 \nu_5 \nu_6 \nu_7}
 & = &
 \sum\limits_{j=j_{\mathrm{min}}}^\infty
 \eps^j \;  I_{\nu_1 \nu_2 \nu_3 \nu_4 \nu_5 \nu_6 \nu_7}^{(j)}
\eq
In this letter we present a method to systematically compute the $j$-th term of the $\eps$-expansion.
The result is expressed in terms of iterated integrals \cite{Chen}.
If $\omega_1$, ..., $\omega_k$ are differential 1-forms on a manifold $M$ and $\gamma : [0,1] \rightarrow M$ a path,
we write for the pull-back of $\omega_j$ to the interval $[0,1]$
\bq
 f_j\left(\lambda\right) d\lambda & = & \gamma^\ast \omega_j.
\eq
The iterated integral is then defined by
\bq
\lefteqn{
 I_{\gamma}\left(\omega_1,...,\omega_k;\lambda\right)
 = } &&
 \nonumber \\
 &&
 \int\limits_0^{\lambda} d\lambda_1 f_1\left(\lambda_1\right)
 \int\limits_0^{\lambda_1} d\lambda_2 f_2\left(\lambda_2\right)
 ...
 \int\limits_0^{\lambda_{k-1}} d\lambda_k f_k\left(\lambda_k\right).
\eq
Multiple polylogarithms are iterated integrals, where all differential one-forms are of the form
\bq
 \omega_j & = & \frac{d\lambda}{\lambda-c_j}.
\eq
If $f(\tau)$ is a modular form, we simply write with a slight abuse of notation $f$ instead of $2 \pi i f d\tau$ in the arguments of iterated
integrals.


\section{The kinematic variables for the multiple polylogarithms}
\label{sect:polylogs}

\begin{figure}
\begin{center}
\includegraphics[scale=1.0]{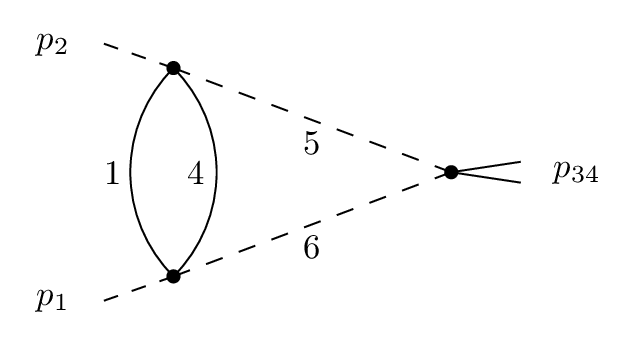}
\end{center}
\caption{
A sub-topology leading to the square root $\sqrt{-s(-4m^2-s)}$.
}
\label{fig_graph_m4}
\end{figure}

The Feynman integral is a function of two kinematic ratios, say $s/m^2$ and $t/m^2$.
A significant fraction of the sub-topologies depends only on $s/m^2$, but not on $t/m^2$.
These integrals are expressible in terms of multiple polylogarithms and their system of differential equations
can be transformed to an $\eps$-form. This introduces square roots, which are absorbed by a change of kinematic variables.
The square root $\sqrt{-s(4m^2-s)}$ is typical for massive Feynman integrals, however there are also sub-topologies, which lead to
the square root $\sqrt{-s(-4m^2-s)}$ (note the sign in front of $4m^2$).
An example is shown in fig.(\ref{fig_graph_m4}).
A transformation, which absorbs both square roots simultaneously is given by
\bq
 \frac{s}{m^2} = - \frac{\left(1+x^2\right)^2}{x\left(1-x^2\right)},
 & &
 \frac{t}{m^2} = y.
\eq
This defines the variables $x$ and $y$. The variable $y$ is not needed for integrals depending only on $s/m^2$.
For the integrals depending only on $s/m^2$ we introduce five differential one-forms
\bq
\label{def_omega}
 \omega_0
 & = &
 \frac{ds}{s} 
 \; = \;  
 \frac{2 dx}{x-i}
 +
 \frac{2 dx}{x+i}
 - \frac{dx}{x-1} - \frac{dx}{x+1} - \frac{dx}{x},
 \nonumber \\
 \omega_4
 & = &
 \frac{ds}{s-4m^2}
 \; = \;
 \frac{2 dx}{x-\left(1+\sqrt{2}\right)}
 +
 \frac{2 dx}{x-\left(1-\sqrt{2}\right)}
 \nonumber \\
 & &
 - \frac{dx}{x-1} - \frac{dx}{x+1} - \frac{dx}{x},
 \nonumber \\
 \omega_{-4}
 & = &
 \frac{ds}{s+4m^2}
 \; = \;
 \frac{2 dx}{x-\left(-1+\sqrt{2}\right)}
 +
 \frac{2 dx}{x-\left(-1-\sqrt{2}\right)}
 \nonumber \\
 & &
 - \frac{dx}{x-1} - \frac{dx}{x+1} - \frac{dx}{x},
 \nonumber \\
 \omega_{0,4}
 & = &
 \frac{ds}{\sqrt{-s\left(4m^2-s\right)}}
 \; = \; 
 \frac{dx}{x-1} - \frac{dx}{x+1} + \frac{dx}{x},
 \nonumber \\
 \omega_{-4,0}
 & = &
 \frac{ds}{\sqrt{-s\left(-4m^2-s\right)}}
 \; = \;
 -\frac{dx}{x-1} + \frac{dx}{x+1} + \frac{dx}{x}.
 \nonumber \\
\eq
Then all sub-topologies, which depend only on $s/m^2$, can be expressed as iterated integrals with letters given
by these five differential one-forms. From eq.~(\ref{def_omega}) it is clear that they are expressible in terms
of multiple polylogarithms.


\section{Elliptic curves}
\label{sect:elliptic_curves}

Let us consider an elliptic curve defined by the quartic equation
\bq
 E
 & : &
 w^2 \; = \; \left(z-z_1\right) \left(z-z_2\right) \left(z-z_3\right) \left(z-z_4\right).
\eq
We set
\bq
 Z_1 & = & \left(z_2-z_1\right)\left(z_4-z_3\right),
 \nonumber \\
 Z_2 & = & \left(z_3-z_2\right)\left(z_4-z_1\right),
 \nonumber \\
 Z_3 & = & \left(z_3-z_1\right)\left(z_4-z_2\right)
\eq 
and define the modulus and the complementary modulus
\bq
 k^2 
 \; = \; 
 \frac{Z_1}{Z_3},
 & &
 \bar{k}^2 
 \; = \;
 \frac{Z_2}{Z_3}.
\eq
Our standard choice for the periods is
\bq
 \psi_1 
 \; = \; 
 \frac{4 K\left(k\right)}{Z_3^{\frac{1}{2}}},
 & &
 \psi_2
 \; = \; 
 \frac{4 i K\left(\bar{k}\right)}{Z_3^{\frac{1}{2}}},
\eq
where $K(x)$ denotes the complete elliptic integral of the first kind.
\begin{figure}
\begin{center}
\includegraphics[scale=1.0]{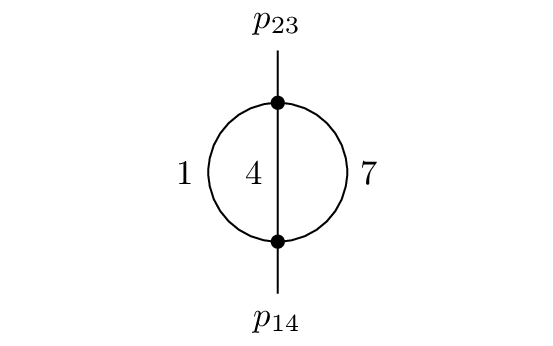}
\includegraphics[scale=1.0]{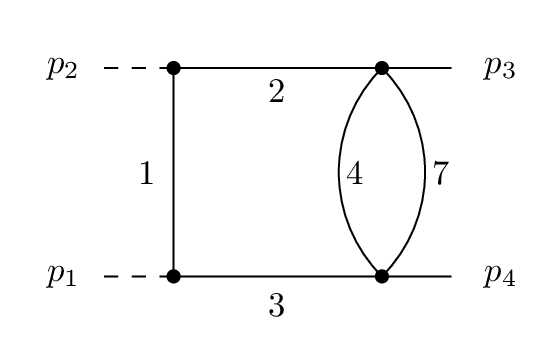}
\includegraphics[scale=1.0]{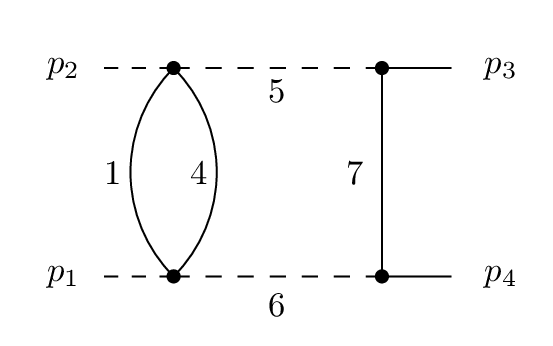}
\end{center}
\caption{
The Feynman graphs associated to the three elliptic curves.
}
\label{fig_elliptic_topologies}
\end{figure}
For the double box integral we have to consider three elliptic curves $E^{\curveone}$, $E^{\curvetwo}$ and $E^{\curvethree}$,
which occur for the first time in the three Feynman graphs shown in fig.~(\ref{fig_elliptic_topologies}).
The equations of the elliptic curves are extracted from the 
maximal cuts of these Feynman integrals \cite{Baikov:1996iu,Lee:2009dh,Kosower:2011ty,CaronHuot:2012ab,Primo:2016ebd,Frellesvig:2017aai,Bosma:2017ens,Harley:2017qut},
specifically from the maximal cuts of
\bq
I_{1001001}\left(2-2\eps\right),
 \;\;
I_{1112001}\left(4-2\eps\right),
 \;\;
I_{2001111}\left(4-2\eps\right).
 \nonumber
\eq
For these three integrals, the elliptic curves are most easily obtained from the loop-by-loop approach in the Baikov representation \cite{Frellesvig:2017aai}. 
We find for all three curves
\bq
 z^{(a,b,c)}_1 \; = \; \frac{t}{m^2}-4,
 \;\;\;
 z^{(a,b,c)}_4 \; = \; \frac{t}{m^2}.
\eq
They differ in the values for the roots $z_2$ and $z_3$.
We have
\bq
 z^{\curveone}_{2,3} & = & -1 \mp 2 \sqrt{\frac{t}{m}},
 \\
 z^{\curvetwo}_{2,3} & = & -1 \mp 2\sqrt{\frac{t}{m^2} + \frac{\left(m^2-t\right)^2}{m^2s}},
 \nonumber \\
 z^{\curvethree}_{2,3} & = & - \frac{\left(s+4t\right)}{\left(s-4m^2\right)} \mp \frac{2}{4m^2-s} \sqrt{\frac{s}{m^2} \left( st + \left(m^2-t\right)^2 \right) }.
 \nonumber
\eq
It is easily checked by computing the $j$-invariants that the three curves are not isomorphic.
However, the curves $E^{\curvetwo}$ and $E^{\curvethree}$ degenerate to curve $E^{\curveone}$ in the limit $s\rightarrow \infty$.
Associated to the curve $E^{\curveone}$ are modular forms of $\Gamma_1(6)$.
We set
\bq
 g_{n,r} & = & - \frac{1}{2} \frac{y\left(y-1\right)\left(y-9\right)}{y-r} \left( \frac{\psi^{\curveone}_1}{\pi} \right)^n,
 \nonumber \\
 p_{n,s} & = & - \frac{1}{2} y\left(y-1\right)^{1+s}\left(y-9\right) \left( \frac{\psi^{\curveone}_1}{\pi} \right)^n,
\eq
Relevant to the problem is the set
\bq
\label{modular_forms}
 \left\{ 1, g_{2,0}, g_{2,1}, g_{2,9}, g_{3,1}, p_{3,0}, g_{4,0}, g_{4,1}, g_{4,9}, p_{4,0}, p_{4,1} \right\}.
\eq
These are modular forms of $\Gamma_1(6)$ in the variable $\tau_6=\psi_2^{\curveone}/(6\psi_1^{\curveone})$, which we may substitute for the variable $y$.


\section{Master integrals and differential equations}
\label{sect:masters}

In order to derive the system of differential equations we first used
{\tt Reduze} \cite{vonManteuffel:2012np},
{\tt Kira} \cite{Maierhoefer:2017hyi} and
{\tt Fire} \cite{Smirnov:2014hma} for the integral reductions.
Taking trivial symmetry relations into account, all programs give 45 master integrals.
However, the reductions disagree for the three most complicated topologies.
For a given set of master integrals $\vec{I}$ we obtain the system of differential equations
\bq
 d \vec{I} & = & A\left(\eps,x,y\right) \vec{I}.
\eq
In general, this system is not yet linear in $\eps$, but it should satisfy the integrability condition
\bq
 d A & = & A \wedge A.
\eq
At  first sight, the results of two of three programs above fail the integrability check.
Still, all three programs correctly implement the 
Laporta algorithm \cite{Laporta:2001dd}.
However, the Laporta algorithm does not guarantee that all relations among the Feynman integrals are found.
\begin{figure}
\begin{center}
\includegraphics[scale=1.0]{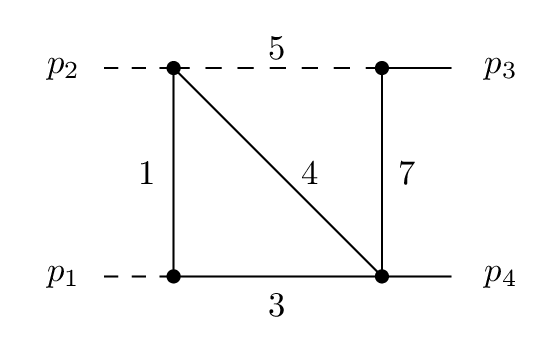}
\end{center}
\caption{
A sub-topology with an additional relation.
}
\label{fig_sector_93}
\end{figure}
Here, we have an example where one additional relation exists in the sub-topology shown in fig.~(\ref{fig_sector_93}).
This additional relation reduces the number of master integrals in this topology from $5$ to $4$.
Imposing this relation, the results from {\tt Reduze}, {\tt Kira} and
{\tt Fire} agree and the integrability condition is satisfied.
In addition, we verified numerically the first few terms in the $\eps$-expansion of this relation.
The extra relation comes from an IBP-identity of a higher sector (i.e. the topology $I_{\nu_1\nu_2 0 \nu_4\nu_5\nu_6\nu_7}$), 
where the coefficients of the integrals of the higher sector are zero.
We would like to add that {\tt Reduze} is able to find the relation and can be forced to use this relation with the command
\verb|distribute_external|\footnote{We thank L. Tancredi for pointing this out and A. von Manteuffel for advice on {\tt Reduze}.}.

In this letter we are interested in the integral $I_{1111111}$. With the help of the methods from \cite{Adams:2017tga}
we may decouple two integrals in the top topology.
Thus we have to consider a system of $42$ master integrals for $I_{1111111}$.

Under a change of basis
\bq
 \vec{J} & = & U \vec{I},
\eq
the differential equations transform into
\bq
 d \vec{J} & = & A' \vec{J},
 \;\;\;
 A' \; = \; U A U^{-1} + U d U^{-1}.
\eq
The main result of this letter is that there exists a transformation $U$, such that
\bq
\label{linear_form}
 d \vec{J}
 & = &
 \left( A^{(0)}\left(x,y\right) + \eps A^{(1)}\left(x,y\right) \right) \vec{J},
\eq
and $A^{(0)}$ is strictly lower triangular
(i.e. $A^{(0)}_{ij}=0$ for $j \ge i$).
The system of differential equations is linear in $\eps$ and easily solved order by order in $\eps$
in terms of iterated integrals.
The transformation matrix is rational in 
\bq
 \eps, x, y, \psi_1^{\curveone}, \psi_1^{\curvetwo}, \psi_1^{\curvethree},
 \partial_y \psi_1^{\curveone}, \partial_y \psi_1^{\curvetwo}, \partial_y \psi_1^{\curvethree}.
\eq
We constructed this matrix by analysing the Picard-Fuchs operators in the diagonal blocks \cite{Adams:2017tga}
and by using a slightly modified version of the algorithm of Meyer \cite{Meyer:2016slj,Meyer:2017joq} for the non-diagonal blocks.
To give an example, the three master integrals in the topology $I_{\nu_1\nu_2\nu_3\nu_400\nu_7}$
can be taken as
\bq
 J_{24}
 & = \;\; & 
 \eps^3 
 \frac{\left(1+x^2\right)^2}{x\left(1-x^2\right)}
 \frac{\pi}{\psi^{\curvetwo}_1} I_{1112001},
 \nonumber \\
 J_{25}
 & = \;\; & 
 \eps^3 \left(1-2\eps\right) 
 \frac{\left(1+x^2\right)^2}{x\left(1-x^2\right)}
 I_{1111001}
 + R_{25,24} \frac{\psi^{\curvetwo}_1}{\pi} J_{24},
 \nonumber \\
 J_{26}
 & = \;\; & 
 \frac{6}{\eps} 
 \frac{\left(\psi^{\curvetwo}_1\right)^2}{2 \pi i W^{\curvetwo}_y} \frac{d}{dy} J_{24}
 + R_{26,24} \left( \frac{\psi^{\curvetwo}_1}{\pi} \right)^2 J_{24}
 \nonumber \\
 & &
 - \frac{\eps^2}{24} \left(y^2-30y-27\right) \frac{\psi^{\curvetwo}_1}{\pi} {\bf D}^- I_{1001001},
\eq
where $R_{25,24}$ and $R_{26,24}$ are rational functions in $(x,y)$, 
${\bf D}^-$ denotes the dimension shift operator $D \rightarrow D-2$
and $W^{\curvetwo}_y$ the Wronskian
\bq
 W^{\curvetwo}_y & = &
 \psi_1^{\curvetwo} \partial_y \psi_2^{\curvetwo}
 -
 \psi_2^{\curvetwo} \partial_y \psi_1^{\curvetwo}.
\eq
As in the sunrise sector \cite{Adams:2018yfj}, one integral is divided by a period ($J_{24}$), while a second integral is given as a derivative plus additional terms ($J_{26}$).
This pattern applies to all elliptic sectors.

The matrix $A^{(0)}$ in eq.~(\ref{linear_form}) vanishes for $x=0$ or $y=1$.
The occurrence of $\eps^0$-terms in the differential equations is expected from the study of the sunrise integral with unequal masses \cite{Adams:2014vja,Bloch:2016izu}.
For $y=1$ the entries of $A^{(1)}$ reduce to the differential one-forms of eq.~(\ref{def_omega}),
for $x=0$ they reduce to the modular forms of eq.~(\ref{modular_forms}).
The solution reduces therefore to multiple polylogarithms for $y=1$ and to iterated integrals of modular forms for $x=0$.
We have compared numerically the solutions of all master integrals with results from  \verb|sector_decomposition| \cite{Bogner:2007cr} and found
agreement.
Albeit the transformation $U$ significantly simplifies the system of differential equations, the length of the solution still exceeds
the format of this letter. 
The definition of the master integrals, the differential equation and the results 
are given in a supplementary electronic file attached to this article.
In addition, a longer publication \cite{Adamsandmore} describes the details of our calculation.

\section{Conclusions}
\label{sec:conclusions}

In this letter we analysed the planar double box integral relevant to top pair production with a closed top loop.
This integral depends on two scales and involves several elliptic sub-sectors.
This integral has not been known analytically and impedes further progress on the analytic computation of higher-loop Feynman integrals with
massive particles.
In this letter we reported that we may transform the system of differential equations to a form linear in $\eps$,
where the $\eps^0$-term is strictly lower-triangular.
With such a linear form the solution in terms of iterated integrals is immediate.
Our techniques open the door for more complicated Feynman integrals.

\subsection*{Acknowledgements}

L.A. and E.C. are grateful for financial support from the research training group GRK 1581.
S.W. would like to thank the Hausdorff Research Institute for Mathematics
for hospitality, where part of this work was carried out.

\bibliography{/home/stefanw/notes/biblio}
\bibliographystyle{/home/stefanw/latex-style/h-physrev5}

\end{document}